\documentclass[11pt]{article}
\usepackage{times}
\usepackage{cospar}
\usepackage[sectionbib]{natbib}
\renewcommand{\bibsection}{\section*{REFERENCES}}
\renewcommand{\bibhang}{\setlength{8mm}}
\renewcommand{\bibsep}{\setlength{0mm}}
\title{PULSAR WIND NEBULAE: THEORETICAL ASPECTS AND OBSERVATIONAL CONSTRAINTS}

\author{Roger A. Chevalier
\address{Department of Astronomy, University of Virginia, P.O. Box 3818, \\
Charlottesville, VA 22903, USA}}

\begin{document}

\def\setoff#1{\noindent\hang\hbox to 40truept{\bf #1\hss}\ignorespaces}
\def\Msun{M_\odot}
\def\la{\raise0.3ex\hbox{$<$}\kern-0.75em{\lower0.65ex\hbox{$\sim$}}}
\def\ga{\raise0.3ex\hbox{$>$}\kern-0.75em{\lower0.65ex\hbox{$\sim$}}}
\newcommand{\kms}{\rm ~km~s^{-1}}
\newcommand{\ergs}{\rm ~erg~s^{-1}}

\maketitle

\begin{abstract}
Of the known pulsar wind nebulae, 8 are good candidates for
being in the early stage of evolution where the wind nebula
is interacting with the freely expanding  supernova ejecta.
Several of these have been identified with historical supernovae.
Although the identification of SN 1181 with 3C 58 has been
thought to be relatively secure, the large size of the nebula, the
amount of swept up mass, and the internal energy indicate a
larger age.
For G11.2--0.3, the nebular size and internal energy are consistent with
the identification with
the possible supernova of 386.
Although the Crab Nebula appears to have approximate energy
equipartition between particles and the magnetic field,
the nebulae 3C 58 and MSH 15--52 appear to be particle dominated.
The low magnetic field is consistent with models in which the nebulae
are created by a shocked pulsar wind.
\end{abstract}


\section*{INTRODUCTION}

Pulsars are expected to be born inside the supernova explosions of massive
stars, which provide the surroundings for the initial evolution
of their wind nebulae \citep{RC84}.
The PWNe (pulsar wind nebulae) initially expand in the freely
expanding ejecta of the supernova.
Eventually, the reverse shock wave from the supernova interaction
with the surrounding medium makes its way back to the center
where it can crush the PWN.
This later phase of evolution has recently been the subject of
detailed studies \citep{vdS01,BCF01}.
In particular, \citet{BCF01} noted that the reverse shock
is likely to be asymmetric so that the PWN can be displaced
from its position over the pulsar.
This scenario provides an explanation for the displacement of the
radio emitting PWN in the Vela remnant \citep{BTG98} and other
remnants.

Here, I emphasize PWNe that are likely to be in the earlier
phase of evolution, before the reverse shock effects.
Recent discoveries at X-ray and radio wavelengths have substantially
increased the number of such objects.
In \S~2, the possible members of this class are listed.
For objects with an approximately constant pulsar power,
the expansion in a supernova is treated in \S~3.
Constraints implied by the energy in the nebulae are discussed
in \S~4.
The conclusions are in \S~5.

\section*{YOUNG PULSAR WIND NEBULAE}

A list of probable young PWNe in which central pulsars have
been identified is given in Table 1; these objects are plausibly
interacting with ejecta.
\begin{table}[here]
\begin{center}
Table 1.  \quad Young pulsars/pulsar wind nebulae

\begin{tabular}{cccccccc}
\hline
\\
Object &   $P$ &  $P/2\dot P$  &  Age  & SN & SNR &  Swept up & Refs. \\
   &  (msec)     &   (year)   &  (year)    &   & & ejecta? & \\
\hline
0540--69 &  50 &   1660    & 760  &   &  Yes &  Yes (optical) & 1 \\
3C 58 &  66 &   5390    & 821  &  1181 & & Yes (X-ray)  &  2  \\
Crab &  33 &   1240    & 948  &  1054  & &   Yes (optical)  & 3 \\
Kes 75 &  325 &   723    &    &   & Yes &  &  4  \\
G292.0+1.8 &  135 &   2890    & $\la 1600$  &  & Yes &   Maybe (optical) & 5,6 \\
G11.2--0.3 &  65 &   24,000    & 1616  & 386  & Yes    &  & 7 \\
MSH 15--52 &  150 &   1700    & 1817  &  185 & Yes &  &  8  \\
G54.1+0.3 &  137 &   2890    &    &   &  &  & 9,10  \\
\hline

\end{tabular}
\end{center}
References: (1) \citet{K89}; (2) \citet{Bet01}; 
(3) \citet{SH97}; 
(4) \citet{H03}; (5) \citet{Cet02b}; (6) \citet{H01}; (7) \citet{R02}; 
(8) \citet{G01}; (9) \citet{L02}; (10) \citet{C02a}

\end{table}
The second column gives the observed pulsar period, $P$, and the
third column gives the characteristic pulsar age, $t_{ch}=P/2\dot P$.
If the pulsar is born rotating much more rapidly than the
current rate and the braking index is $n=3$, then $t_{ch}$
is the actual age.
If the pulsar is born with a period close to its current
period, it can be younger than $t_{ch}$.
Alternatively, if the pulsar is born spinning rapidly and
has a braking index $n<3$, it can be older than $t_{ch}$.
The fourth column is an estimate of the actual age.
For 0540--69 \citep{K89} and G292.0+1.8 \citep{MC79},
the age estimate is from the size of the nebula and velocities
found from optical spectroscopy.
For the other objects with ages, the estimate is from a tentative
supernova identification, given in the next column.
The identification of the Crab with SN 1054 is generally considered very
secure, but the other identifications are less secure.
There is still some uncertainty over whether all of the events are
in fact supernovae, e.g.,  SN 386 \citep{SG02}.
The next column indicates whether an extended supernova remnant
is observed around the PWN, and the penultimate 
column indicates whether there is 
evidence for ejecta swept up by the PWN.

There are indications that these nebulae are in the early
phase of interaction with freely expanding ejecta.
One expectation in this picture is that the PWN should
shock and sweep up a shell of supernova ejecta \citep{RC84,CF92}.
The shock wave is initially expected to be radiative, but it
becomes nonradiative due to the  decline in the supernova density.
A good example of an ejecta shell is the complex of  filaments in the
Crab Nebula.
In this case, there is evidence for the shock wave in the ejecta
\citep{SH97};
it may be in the process of making a transition from radiative
to nonradiative.
For 3C 58, there is evidence for X-ray emission from swept-up
ejecta \citep{Bet01}, implying the presence of a
nonradiative shock.
The optical emission from 0540--69 may be from a radiative shock \citep{K89,CF92}.
The optical emission from G292.0+1.8 appears to be from the vicinity of
the PWN \citep{MC79,H01}, but the relation between them has not yet been determined.

Another expectation of the model with expansion in ejecta is that the
pulsar should be centrally located within the PWN.
This appears to be true for the objects listed in Table 1, although in
most cases the PWN has an asymmetric boundary.
The PWN in G292.0+1.8 is substantially off the center of the surrounding
supernova remnant, which has led to the suggestion that the pulsar
has a velocity $\sim 770\kms$ \citep{H01}.
If this is the case, the pulsar is expected to move to a place in the
ejecta where it is comoving with the surrounding gas and it is surrounded
by uniformly expanding ejecta.
Some degree of asymmetry may develop if there is a gradient in the
surrounding density distribution.

\section*{MODELS FOR 3C 58 AND G11.2--0.3}

If we assume that 3C 58 and G11.2--0.3 are associated with SN 1181
and SN 386, respectively, models for the PWNe can make use of
the fact that the characteristic age is much larger than the 
true age, so that the pulsars have not significantly spun down.
This allows the assumption that the pulsar power, $\dot E$, is
constant during the evolution.

We begin by considering the expansion of the PWN into the
inner parts of the supernova ejecta.
The density distribution into which the PWN initially expands
can be estimated from explosion models.
\citet{CF92} used simple power law models for the density distribution.
More detailed models have been considered by
\citet{MM99}, who give asymptotic forms for the
the final density distribution at low and high velocities.
For the cases to be studied here, the asymptotic low velocity
density profile is applicable over the time of interest.
For an explosion in a star with a radiative envelope, the inner density
profile, using eq. (46) of \citet{MM99}, can be expressed as
\begin{equation}
\rho t^3 = 4.3\times 10^8\left(v\over 1000\kms\right)^{-1.06}
\left(M_{ej}\over 10~\Msun\right)^{1.97} 
 E_{51}^{-0.97} {\rm~g~cm^{-3}~s^3},
\end{equation}
where
$v=r/t$ is the free expansion velocity,
$M_{ej}$ is the total ejected mass, and
$E_{51}$ is the explosion energy in units of $10^{51}$ ergs.
For a star with a convective envelope, the density distribution may
be flatter, but the density at $1000\kms$ is close to the above value.

The expansion of a PWN in a density distribution can be approximately
treated as a thin shell driven by a uniform pressure wind bubble
with adiabatic index $\gamma=4/3$ \citep{OG71,C77,RC84}.
The radius can be found analytically if the pulsar power, $\dot E$, is
constant and the surrounding medium has a power law density distribution.
As discussed above, this may be the case for 3C58 and G11.2--0.3.
Then, from eq. (2.6) of \citet{CF92}, we have
\begin{equation}
R_p=0.59 \dot E_{38}^{0.254}E_{51}^{0.246}
\left(M_{ej}\over 10\Msun\right)^{-0.50}
\left(t\over 10^3 {\rm~yr}\right)^{1.254} {\rm~pc},
\end{equation}
where $\dot E_{38}$ is $\dot E$ in units of $10^{38}\ergs$.
In cases where $R_p$, $\dot E$, and $t$ are known, we can
solve for $M_{ej}/E_{51}^{0.49}$;
because $E_{51}\sim 1$ is expected, we have an estimate of the
total ejecta mass.
These quantities are known for 3C 58 and G11.2--0.3 if the historical
supernova identifications are assumed, and the results are
given in Table 2.
These estimates involve a spherical approximation for PWNe
that are apparently not spherical, but an average radius can be taken
 \citep{W97,R02}.
 The assumed distances are 3.2 kpc for 3C 58 \citep{R93} and
 5 kpc for G11.2--0.3 \citep{G88}.
The expected values of $M_{ej}$ for a core collapse supernova are
typically several $\Msun$ or more.
In the case of a very fast Type Ic supernova, SN 1994I, $M_{ej}$ was only
$\sim 1\Msun$ \citep{I94}, which is an extreme case.
The point is that the model leads to a value of $M_{ej}$ that is
smaller than expected for 3C 58.
The radius is larger than would be expected if it were expanding into a
normal supernova.
There is no problem with G11.2--0.3.

\begin{table}[h]
\begin{center}
Table 2. \quad Estimated ejecta  and swept up mass

\begin{tabular}{cccccc}
\hline
\\
Object &  $\dot E_{38}$ &  Radio radius&  $M_{ej}/E_{51}^{0.49}$ &  Predicted  &  Observed  \\
      &          &   (pc)  &  ($\Msun$)   & $M_{sw}$ ($\Msun$)    &  $M_{sw}$ ($\Msun$)    \\
\hline
3C 58 &  0.27 &   3.3    & 0.1&  0.002 &   0.1  \\
G11.2--0.3 &  0.064 &   0.9    & 3.5 &  0.05 & \\
\hline
\end{tabular}
\end{center}
\end{table}

Another constraint comes from the amount of mass swept up by
the wind bubble, $M_{sw}$.
An integration over the central density shows that
$M_{sw}\approx 1.0 \dot E R_p^{-2}  t^3$,
fairly independent of supernova density distribution.
Values of $M_{sw}$ deduced in this way for 3C 58 and G11.2--0.3 are
given in Table 2.
The value of $0.002\Msun$ deduced for 3C 58 can be compared to the $0.1\Msun$
found from X-ray observations \citep{Bet01}.
Again, there is a problem with the model mass being too low.
The mass could be brought into agreement with the observed
value if the age were increased by a factor $\sim 3$.

These models assume that the supernova ejecta are swept into
a thin shell that remains spherically symmetric.
However, the shell is being accelerated and is subjected to
Rayleigh-Taylor instabilities, which can decrease the coupling
between the pulsar wind bubble and the swept up gas.
In the limit that there is no further acceleration after
the ejecta are shocked, the PWN radius (eq. [2]) is increased
by a factor of 1.4.
The masses $M_{ej}$ and $M_{sw}$ are increased by a factor of 2,
which does not change the conclusion about the difficulties with 3C 58.

The model with constant $\dot E$ can be used to predict the
internal energy in the PWN.
This energy is reduced from the total deposited energy, $\dot E t$,
because of work done on the surrounding supernova gas.
For a range of flat central density distributions, the
internal energy is  $0.45\dot E t$ (Table 1 of
Chevalier and Fransson 1992).   
The internal energy in a PWN has relativistic particle and magnetic
field components;
a minimum value for the total energy in
particles and fields can be found from the synchrotron luminosity
and the emitting volume (e.g., Pacholczyk 1970).  
\citet{T02} discuss the radio emission from the PWN in G11.2--0.3.
The value of the minimum energy deduced from the radio emission
for the two PWNe is given in Table 3.
The actual energy must be larger when a larger frequency range
is considered.
It can be seen that there is not a problem with the energy for G11.2--0.3,
but that the energy in 3C 58 appears to be larger than that expected
for the observed pulsar and the designated age.
A larger age for 3C 58 would allow a larger energy to be deposited in
the nebula.

\begin{table}[here]
\begin{center}
Table 3. \quad Minimum energy in radio emitting particles ($10^7-10^{11}$ Hz)
\begin{tabular}{ccccc}
\hline
\\
Object & $L_{radio}$ & $\dot E t$ &  $E_{min}$   &  $E_{min}/\dot E t$  \\
      & ($10^{34}$ ergs s$^{-1}$) & ($10^{48}$ ergs) & ($10^{48}$ ergs)             &    \\
\hline
3C 58 & 2.8  &     0.7 &   1.0    & 1.5 \\
G11.2--0.3 & 0.12  & 0.3  &   0.03    & 0.1 \\
\hline
\end{tabular}
\end{center}
\end{table}%

There are thus several arguments for 3C 58 being older than SN 1181,
even though \citet{SG02} consider the identification to be secure.
The problems are that the  PWN is too large to be expanding into a normal supernova,
the expected mass swept up by PWN smaller than observed,
and the internal energy is larger than can be supplied by the pulsar.
A larger age for the remnant is consistent with the 
slow expansion of 3C 58 observed
at both radio \citep{BKW01} and
optical \citep{FKB88}
wavelengths.

\section*{ENERGY EQUIPARTITION IN PWNe}

One of the important properties of a PWN is the relative amount of
energy in particles and in magnetic fields.
In the Crab Nebula, there is approximate equipartition overall
in these energies.
In the detailed MHD model of \citet{KC84} for the Crab Nebula,
this property is produced by the choice of the $\sigma$ parameter,
the ratio of Poynting flux to particle kinetic energy flux in the
pulsar wind.
The value $\sigma\approx 0.003$ is deduced; the magnetic field is
relatively weak in the wind and is increased by the shock compression
and further compression in the decelerating postshock flow.
This value of $\sigma$ is close to the upper limit that is allowed in
this kind of model, or the flow would not be able to decelerate to
meet the outer boundary condition, but there is no particular
reason for   this value to be produced.

One way to estimate the overall magnetic field in a PWN is from
the synchrotron break frequency, $\nu_{br}$,
and the age of the PWN. 
The determination of $\nu_{br}$ depends on the spectrum of the nebula.
There is increasing evidence that the particle spectrum injected
into PWNe typically has at least one intrinsic spectral break.
Models for the radio to X-ray emission with a single power law injection
spectrum generally fail \citep{RC84}, and the well-observed Crab
Nebula spectrum requires an injection spectrum with a break
(e.g., Amato et al., 2000).   
With synchrotron losses, the spectrum
develops a further break.

Two PWNe for which there is age information and information
on $\nu_{br}$ are 3C 58 and MSH 15--52.
As discussed above, 3C 58 may be older than 821 years, which I consider
as a lower limit; a lower limit on the age yields an upper limit on the
magnetic field.
MSH 15--52 has been suggested to be the remnant of SN 185, although
 G315.4--2.3 is another
candidate for the remnant of this supernova \citep{SG02}.
However, the large size of the nebula associated with MSH 15--52
\citep{G01} indicates that it is not significantly younger than 1700 years.

The determination of $\nu_{br}$ depends on the interpretation of
the overall spectrum of the PWN.
For 3C 58 and MSH 15--52, as for most PWNe, there are detections
at only radio and X-ray wavelengths.
However, the fact that the extent of the X-ray emission is comparable
to that of the radio emission in both cases indicates that $\nu_{br}$
is not much lower than X-ray energies, assuming that particles originate
close to the pulsar and move out in the nebula.
In the case of 3C 58, the X-ray spectrum steepens at large radii,
showing that synchrotron losses are significant in the X-ray regime \citep{T00}.
An estimate of $h\nu_{br}$ is thus $\sim 0.5$ keV, which is consistent
with the fact that the X-ray spectrum is somewhat steeper than
the spectrum from radio to X-ray wavelengths.
The implications for the magnetic field and magnetic energy
are given in Table 4.
The magnetic field strength in MSH 15--52 can be estimated from
similar arguments \citep{G01} and is given in Table 4.
In this scenario, the low frequency breaks are intrinsic to the
injected particle spectrum; this is a controversial point and there
have been discussions of the low frequency 
breaks in terms of synchrotron losses
(e.g., Woltjer et al., 1997 on 3C 58 and Roberts et al., 2002 on G11.2--0.3).

\begin{table}[here]
\begin{center}

Table 4. \quad Comparison of magnetic and minimum internal energies
\begin{tabular}{ccccc}
\hline
\\
Object & $h\nu_{br}$  &  $B$  & $E_B$ &  $E_{min}$   \\
      & (keV)   &  ($\mu$G)  & ($10^{47}$ ergs) & ($10^{47}$ ergs)               \\
\hline
3C 58 &      0.5 &   16    & 0.4  &  10 \\
MSH 15--52 &  1 &   8   & 0.3  &  7 \\
\hline
\end{tabular}
\end{center}

\end{table}%

It is also possible to 
estimate minimum magnetic plus particle energy, $E_{min}$, from
the radio synchrotron emission ($10^7-10^{11}$ Hz).
The results, given in Table 4, show that the magnetic energy
is considerably less than the total internal energy in both
cases, so that the nebulae are particle dominated.
In the context of the \citet{KC84} model, this would require
a remarkably low Poynting energy flux in the pulsar wind.
However, the result is  consistent with the finding in
the \citet{KC84} model that a particle dominated wind is
needed to produce a shock front and deceleration of the flow
to match the outer boundary.
This suggests that pulsar winds may have a range of magnetizations,
giving rise to a range of nebular properties.
If cases with a highly magnetized wind occur, they would give
rise to something other than a standard pulsar wind nebula.
In this limit, the wind termination shock moves in to the
pulsar \citep{RG74,EC87} so that the immediate surroundings of
the pulsar are in communication with the ambient medium.

\section{DISCUSSION AND CONCLUSIONS}

The increasing number of PWNe are providing many examples which
can be compared to models initially developed for the Crab Nebula.
In a model with interaction with freely expanding supernova ejecta,
the PWN properties can provide a check on the age estimate
for the nebula.
Such models have previously been developed for the Crab Nebula,
MSH 15-52, and 0540--69 \citep{CF92}.
Recent observational results on 3C 58 and G11.2--0.3 allow similar
models to be considered for these objects, which are of special
interest because of their possible identifications with historical supernovae.
The models indicate that 3C 58 is older than SN 1181, but that
G11.2--0.3 is consistent with being the remnant of SN 386.
These tentative conclusions need to be followed up by more detailed
studies of the remnants.
In both cases, the model predicts that the PWN is driving a shock front
into freely expanding ejecta.
Gas shocked in this way may have been observed in 3C 58 \citep{Bet01},
but further observations are needed.
Such gas has not yet been observed in G11.2--0.3.
The external supernova remnant interaction also provides constraints
on the system.
This interaction is clearly seen in G11.2--0.3 \citep{R02}, but
not in 3C 58, implying interaction with low density surroundings
for that case.

Another finding here is that the 3C 58 and MSH 15-52 PWNe are
particle dominated, which requires that the  pulsar
winds have a very low magnetization parameter.
This property may be one of the reasons why these PWNe have
low efficiencies of X-ray luminosity production compared to
the pulsar power \citep{C00}.
3C 58, G292.0+1.8, G11.2--0.3, MSH 15-52, and G54.1+0.3 all
have significantly lower levels of X-ray luminosity production efficiency
than the Crab Nebula.

\section*{ACKNOWLEDGEMENTS}

I am grateful to the referee and editor for helpful comments.
This work was supported in part by NASA grants NAG5-8130 and NAG5-13272.

\smallskip
\noindent
email address for R. A. Chevalier: rac5x@virginia.edu

\end{document}